\documentclass{svproc}
\usepackage{url}

\usepackage{amssymb}
\usepackage{amsmath}
\usepackage{graphicx}
\usepackage{tabularx}
\usepackage{dcolumn}
\usepackage{multicol}
\usepackage{siunitx}
\usepackage{footmisc}
\usepackage{tikz}
\usepackage{pgfplots}
\pgfplotsset{compat = 1.3}
\usepackage{subcaption}

\begin{document}
\mainmatter              
\title{Comparison of stochastic BGK and FP methods for the simulation of non-equilibrium multi-species molecular gas flows}
\titlerunning{Stochastic BGK and FP methods for non-equilibrium multi-species flows}
%
\author{Franziska Hild \and Marcel Pfeiffer}
\authorrunning{Franziska Hild \and Marcel Pfeiffer} 
%
%
\institute{Institute of Space Systems, University of Stuttgart, \\
Pfaffenwaldring 29,  70569 Stuttgart, Germany,\\
\email{hildf@irs.uni-stuttgart.de}}

\maketitle              

\begin{abstract}
Due to limited possibilities of experimental investigations for non-equilibrium gas flows, numerical results are of highest interest.
Although the well-established Direct Simulation Monte Carlo (DSMC) method achieves highly accurate solutions, the computational requirements increase excessively for lower Knudsen regimes.
Computationally more efficient simulations can be achieved with stochastic continuum-based methods using either the Bhatnagar-Gross-Krook (BGK) or the Fokker-Planck (FP) approximations where, instead of particle collisions, particle relaxation processes are considered.
This paper explains the implementation of different stochastic BGK and FP methods in the open-source particle code PICLas for multi-species molecular gas flows.
For verification, the results of different test cases are compared.
\end{abstract}
\section{Introduction}
Looking at multi-scale flow problems across a wide range of Knudsen numbers, the well-established Direct Simulation Monte Carlo (DSMC) method~\cite{bird} achieves highly accurate solutions.
However, the computational requirements increase excessively for lower Knudsen regimes, making it computationally infeasible for many practical applications.
A possible solution is a coupling of the DSMC method with continuum-based methods, leading to reduced computational costs.
Using the same particle approach for simplicity of the coupling, different stochastic particle-based methods such as the Bhatnagar-Gross-Krook (BGK)~\cite{bgk,esbgk,shakhov} and the Fokker-Planck (FP)~\cite{fp-basic,fp,esfp,cubicfp,fp-kim,fp-hepp} approximations for the Boltzmann collision integral can be used for this purpose.
The advantage of these methods in the computational efficiency in transition and continuum regimes, as they consider relaxation processes of the particles instead of binary particle collisions.

In this paper, the implementation of different stochastic BGK and FP methods for multi-species molecular gas flows in the open-source particle code PICLas~\cite{piclas,piclas-bgk} is explained.
First, the different models are presented, including an Ellipsoidal Statistical BGK (ESBGK) model~\cite{esbgk}, a Shakhov BGK model~\cite{shakhov}, and an Ellipsoidal Statistical Fokker-Planck (ESFP) model~\cite{esfp}.
The focus of this paper lies in the differences between the models and their implementation in PICLas, mainly regarding the relaxation and sampling of the particle velocities.
The relaxation of internal energies are described through Landau-Teller equations as shown in Hild et al.~\cite{bgk-polymulti}, which is the same for all models and thus is not presented any further here.
The energy and momentum conservation scheme is discussed briefly, which is crucial for the stability of the simulations.
To assess the accuracy of the different models, the results of two supersonic Couette flow test cases are compared to DSMC reference solutions.

\section{BGK method} \label{s:bgk}
The BGK operator approximates the Boltzmann collision integral by a relaxation process of a particle distribution function $f_s \left(\mathbf x,\mathbf v, t \right)$ of a species $s$ at position $\mathbf x$ and with velocity $\mathbf v$ towards a target distribution $f_s^t$~\cite{bgk}, using the relaxation frequency $\nu$:
\begin{equation}
	\partial_t f_s + \mathbf {v} \partial_{\mathbf x} f_s = \nu(f_s^t-f_s). \label{eq:bgk}
\end{equation}
For atomic species, the target distribution function only consists of a translational part $f_s^{t, \text{tr}}(\mathbf v)$, whereas for molecular species, the distribution function can be separated into a translational, a rotational, and a vibrational part as demonstrated in~\cite{MATHIAUD202265,dauvois2021bgk}.
Both the ESBGK~\cite{esbgk} and the SBGK~\cite{shakhov} models use one relaxation term per species $s$, produce the Maxwellian distribution in the equilibrium state, and fulfill the indifferentiability principle.
While there the H theorem is proven for the ESBGK model, there is no general proof for the SBGK model.
Furthermore, the positivity of SBGK target distribution function $f^\text{SBGK,tr}_s$ cannot be guaranteed.

\subsection{ESBGK mixture model} \label{s:esbgk}
The ESBGK model presented here combines different models of Mathiaud et al.~\cite{MATHIAUD202265}, Pfeiffer~\cite{energy-cons} and Brull~\cite{brull,brull2021ellipsoidal} into a mixture model that allows for non-equilibrium effects in the internal degrees of freedom.
It is already discussed in greater detail in~\cite{bgk-polymulti}.
Instead of a Maxwellian target distribution, the correct Prandtl number is achieved by using an anisotropic Gaussian distribution~\cite{esbgk}:
\begin{equation}
	f_s^{\text{ESBGK,tr}} = \frac{n_s}{\sqrt{\det 2\pi\mathbf A_s}} \exp\left[-\frac{1}{2}\mathbf c^\text{T} \mathbf A_s^{-1} \mathbf c\right]. \label{eq:esbgk}
\end{equation}
The anisotropic matrix $\mathbf A_s$ is defined as
\begin{equation}
	\mathbf A_s = \frac{k_\text{B} T_{\text{tr,rel}}}{m_s} \mathbf I - \frac{1-\alpha Pr}{\alpha Pr}\left(\frac{T_{\text{tr,rel}}}{T_{\text{tr}}} \mathbf\Theta-\frac{k_\text{B}}{m_s}T_{\text{tr,rel}}\mathbf I\right)
\end{equation}
with the identity matrix $\mathbf I$, and the symmetric stress tensor $\mathbf\Theta$, calculated with the pressure tensor $\mathbf P$, and the mass density $\rho$:
\begin{equation}
	\mathbf\Theta = \frac{\mathbf{P}}{\rho} = \frac{1}{\rho} \sum_{s=1}^M m_s\int \mathbf c \mathbf c^\text{T} f_s\,\text d\mathbf v
\end{equation}
The parameter $\alpha$ is a model variable that depends on mass fraction, density fraction and internal degrees of freedom~\cite{brull2021ellipsoidal}:
\begin{equation}
	\alpha = m \frac{\sum_{s=1}^M \frac{n_s}{m_s} \left(5 + \xi_{\text{vib},s}+\xi_{\text{rot},s}\right)}{\sum_{s=1}^M n_s \left(5 + \xi_{\text{vib},s}+\xi_{\text{rot},s}\right)}
\end{equation}
The relaxation frequency $\nu_\text{ESBGK}$ of the model is defined as
\begin{equation}
	\nu_\text{ESBGK} = \frac{nk_{\text{B}} T_{\text{tr}}}{\mu}\alpha Pr
\end{equation}
with the viscosity of the mixture $\mu$ and the translational temperature $T_{\text{tr}}$.
$Pr$ is the targeted Prandtl number of the gas mixture, which is calculated using collision integrals~\cite{collint}.
For molecular species, the rotational and vibrational components of the distribution function are taken directly from~\cite{MATHIAUD202265}.
The relaxation temperatures are chosen to match the Landau-Teller relaxation.
For more details, the reader is referred to~\cite{bgk-polymulti}.

\subsection{Shakhov BGK mixture model}
In the SBGK model, the heat flux is modified to produce the correct Prandtl number~\cite{shakhov}.
Thus, the target distribution function for a mixture model reads
\begin{alignat}{1}
    f^\text{SBGK,tr}_s = f^\text{M}_s &\left[ 1+ \left(1-\alpha Pr\right) \frac{\mathbf{cq}}{5n \left(k_\text{B}T_{\text{tr}}/m_s \right)^2} \left(\frac{\left|\mathbf{c}\right|^2}{2 k_\text{B}T_{\text{tr}}/m_s} - \frac{5}{2}\right) \right] \label{eq:sbgk}
\end{alignat}
with the heat flux vector
\begin{equation}
	\mathbf q =\sum_{s=1}^M m_s\int \mathbf c \left|\mathbf c\right|^2 f_s\,\text d\mathbf v
\end{equation}
and the Maxwellian distribution $f^\text{M}_s$.
The relaxation frequency $\nu_\text{SBGK}$ of the model is defined as
\begin{equation}
	\nu_\text{SBGK} = \frac{nk_{\text{B}} T_{\text{tr}}}{\mu}.
\end{equation}
For molecular species, the rotational and vibrational components of the distribution function are defined similar as for the ESBGK model (see Section~\ref{s:esbgk}).

\section{FP method}
The FP operator approximates the Boltzmann collision term by using a drift vector $\mathbf{a}=a_i$ and a diffusion matrix $\mathbf{D}_s=D_{ij,s}$~\cite{fp-basic}:
\begin{equation}
	\partial_t f_s + \mathbf {v} \partial_{\mathbf x} f_s = -\sum_{i=1}^{3} \frac{\partial}{\partial v_i} (a_i f_s) + \sum_{i=1}^{3}\sum_{j=1}^{3} \frac{\partial^2}{\partial v_i \partial v_j} (D_{ij,s} f_s). \label{eq:fp}
\end{equation}
For molecular species, the internal energies are handled following the BGK approach presented in Section~\ref{s:bgk}.
In this paper, the ESFP model~\cite{esfp} for molecular gas mixtures is presented and compared to the BGK models.
Same as the ESBGK model, it fulfills the H theorem.

\subsection{ESFP mixture model}
The ESFP mixture model modifies the diffusion matrix $\mathbf{D}_s$ to correct the Prandtl number of the Standard FP model~\cite{esfp}:
\begin{equation}
	\mathbf{D}_s^\text{ESFP} = \nu_\text{ESFP} \left( (1-\chi_s) \frac{k_\text{B}T_\text{tr}}{m_s} \mathbf{I} + \chi_s \mathbf\Theta \right).
\end{equation}
$\mathbf{D}_s$ is a convex combination of the stress tensor $\mathbf\Theta$ and its equilibrium value $(k_\text{B}T_\text{tr}/m_s) \mathbf{I}$ for a species $s$ with a parameter $\chi_s$ including $\lambda_\text{max}$ as the (positive) maximum eigenvalue of $\mathbf\Theta$, so that $\mathbf{D}_s$ remains strictly definite positive:
\begin{equation}
	\chi_s = \text{max} \left( 1-\frac{3}{2Pr}, - \frac{k_\text{B}T_\text{tr}/m_s}{\lambda_\text{max} - k_\text{B}T_\text{tr}/m_s} \right).
\end{equation}
The drift vector is defined similar to the Standard FP model:
\begin{equation}
	\mathbf{a}^\text{ESFP} = -\nu_\text{ESFP} \mathbf{c}.
\end{equation}
The relaxation frequency $\nu_\text{ESFP}$ of the model is defined as
\begin{equation}
	\nu_\text{ESFP} = \frac{n k_\text{B} T_\text{tr}}{3\mu} Pr.
\end{equation}

\section{Implementation}
All the particle methods presented in this paper are implemented in the open-source code PICLas~\cite{piclas}.
The different implementations of the models for the translational relaxation of the particles and the sampling of the new velocities are discussed.
In addition, the basic energy and momentum conservation scheme is described (for details see~\cite{bgk-polymulti}).
For details on the rotational and vibrational relaxations, the reader is also referred to~\cite{bgk-polymulti}.

\subsection{Relaxation and sampling of particle velocities} \label{s:relaxation}
For the BGK models, all particles in a cell relax with a probability $P$ in each time step $\mathrm{\Delta} t$ towards the specified target distribution function $f_s^\text{ESBGK,tr}$ or $f_s^\text{SBGK,tr}$:
\begin{equation}
	P = 1-\exp\left[-\nu \mathrm{\Delta} t\right], \quad \nu=\nu_\text{ESBGK},\nu_\text{SBGK}.
\end{equation}
The relaxation frequency $\nu$ is calculated in each cell and time step using collision integrals~\cite{collint}, as described in~\cite{bgk-polymulti}.
If a uniform random number is smaller than the probability, the particle considered relaxes and its new velocity is sampled from the target distribution function of the used method.

For the ESBGK method, the transformation matrix $\mathbf{S}_s$~\cite{gallis-torczynski} with $\mathbf{A}_s=\mathbf{S}_s \mathbf{S}_s^\text{T}$ is approximated and used to transform a random Maxwellian vector $\mathbf{r}_p$ for a particle $p$ to a velocity vector from the ESBGK target distribution:
\begin{align}
	\mathbf{v}_{p,\text{ESBGK}}^* &= \mathbf{u} + \mathbf{S}_s \mathbf{r}_p, \\
	\mathbf{S}_s &= \sqrt{\frac{k_\text{B} T_\text{tr,rel}}{m_s}} \left[ \mathbf{I} - \frac{1-\alpha Pr}{2\alpha Pr}  \left( \frac{m}{k_\text{B} T_\text{tr}} \mathbf\Theta - \mathbf{I} \right) \right].
\end{align}
The superscript $^*$ marks a parameter after relaxation, but before energy and momentum conservation.

For the SBGK method, however, an analytical expression for a similar vector transformation is not available, which is why an acceptance-rejection (AR) algorithm~\cite{accept-reject} is used.
Hereby, an envelope function $g_s(\mathbf{r}_p)$ is defined.
For the presented particle SBGK method, good results are found by choosing
\begin{align}
	g_s(\mathbf{r}_p) &= A_s f_s^\text{SBGK,tr}(\mathbf{r}_p) \\
	A_s &= 1+ 10 \frac{\max (\mathbf{q})}{(k_\text{B}T_\mathrm{tr}/m_s)^{3/2}}
\end{align}
as envelope function and
\begin{equation}
	P_s^\text{SBGK,AR}(\mathbf{r}_p) = 1 + (1-Pr) \frac{\mathbf{r}_p \mathbf q^\text{T}}{5 n (k_\text{B}T_\mathrm{tr}/m_s)^{3/2}} \left( \frac{\left|\mathbf{r}_p\right|^2}{2} - \frac{5}{2} \right)
\end{equation}
as acceptance probability.
A normal distributed vector $\mathbf{r}_p$ is accepted if a uniform random number $r < P_s^\text{SBGK,AR}(\mathbf{r}_p) / A_s$.
The new velocity is then calculated with
\begin{equation}
	\mathbf{v}_{p,\text{SBGK}}^* = \mathbf{r}_p \sqrt{\frac{k_\text{B}T_\text{tr,rel}}{m_s}}.
\end{equation}

For a detailed description of the different BGK sampling options, the reader is referred to~\cite{piclas-bgk}, in which also alternative methods for sampling are described.

The ESFP model relaxes each particle according to the Ornstein-Uhlenbeck process~\cite{ornstein-uhlenbeck} with the following stochastic differential equation to solve the corresponding FP equation~\eqref{eq:fp}:
\begin{equation}
	\text{d}\mathbf{c} = \mathbf{a} \: \text{d}t + \sqrt{2} \mathbf{D}^{1/2} \: \text{d} \mathbf{w}
\end{equation}
$\text{d} \mathbf{w}$ is the standard three-dimensional Wiener process~\cite{esfp}.
Using an exact time integration~\cite{fp-basic}, the new particle velocity for the ESFP model is calculated by
\begin{align}
	\mathbf{v}_{p,\text{ESFP}}^* &= \mathbf{u} + \mathbf{c}_p \exp \left(-\nu_\text{ESFP} \mathrm{\Delta} t \right) + \frac{\left(\mathbf{D}_s^{\text{ESFP}}\right)^{1/2}}{\sqrt{\nu_\text{ESFP}}} \sqrt{1 - \exp \left( -2\nu_\text{ESFP} \mathrm{\Delta} t \right)} \cdot \mathbf{r}_p
\end{align}
with $\mathbf{c}_p$ being the thermal particle velocity before relaxation.

\subsection{Energy and momentum conservation}
Due to the stochastic approach with particles, an energy and momentum conservation scheme needs to be established even though the presented BGK and FP models are in general both momentum and energy conserving.
The reason for this lies in the random choice of the new states of the particles
from the distribution functions as described in Section~\ref{s:relaxation}, whereby the conservation of energy and momentum is not automatically given.
Thus, a large number of particles would be needed for stable simulation results if no additional energy and momentum conservation step would be implemented.
In general, the chosen conservation scheme is based on the work in~\cite{energy-cons,burt-boyd} and extended to polyatomic molecules and mixtures of different gas species in~\cite{bgk-polymulti}.
The main idea is to distribute the energy difference between the old energy and the new energy from the sampling evenly according to the respective degrees of freedom to the translational energy of all particles in the cell and the internal energies of the respective relaxing particles within the cell~\cite{energy-cons}.
More details on the implementation can be found in~\cite{bgk-polymulti}.

\section{Simulation results for supersonic Couette flows} \label{sec:couette}
The different multi-species BGK and FP implementations are verified and compared with a test case of two one-dimensional supersonic Couette flows.
Simulating a 50\%-50\% N$_2$-He mixture first, the initial particle density is $n_0=\SI{1.3e20}{\per\cubic\meter}$, leading to a Knudsen number of $Kn_{\text{VHS}}=0.0112$ with a characteristic length of $L=\SI{1}{\meter}$.
As second case, an air mixture with N, O, N$_2$, O$_2$, and NO, with $n_0=\SI{1.25e20}{\per\cubic\meter}$ and a corresponding Knudsen number of $Kn_{\text{VHS}}=0.0126$ is simulated in order to test the models with a more complicated case of multiple molecular species in a mixture.
The velocity and temperature of the gas are initialized at $v_0=\SI{0}{\meter\per\second}$ and $T_0=\SI{273}{K}$ in both cases.
The boundaries in $y$ direction have a velocity of $v_\text{wall,1}=\SI{350}{\meter\per\second}$ and $v_\text{wall,2}=\SI{-350}{\meter\per\second}$, respectively, assuming diffuse reflection and complete thermal accommodation at a constant wall temperature of $T_\text{wall}=\SI{273}{K}$.
The time steps and particle weighting factors are chosen so that the mean free path and the collision frequency are resolved.
The VHS model is used~\cite{bird}, and the parameters for a collision pair of unlike species are determined as an average (referred to as collision-averaged).

\begin{figure}
	\centering
	\subfloat[N$_2$-He mixture]{\includegraphics{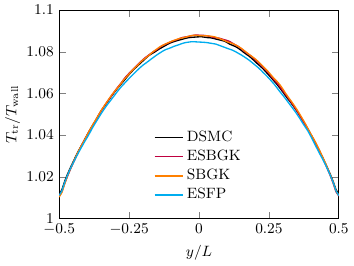}}
	\subfloat[Air mixture]{\includegraphics{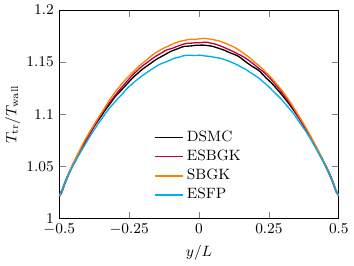}}
	\caption{Translational temperature for the ESBGK (purple), SBGK (orange), and ESFP (blue) models as well as the DSMC result (black) as reference.}
	\label{fig:couette}
\end{figure}

The translational temperature results in comparison for the different methods are shown in Figure~\ref{fig:couette}.
While there is very good agreement for the ESBGK and SBGK models compared to the DSMC reference solution, the ESFP model shows deviations in the peak temperature.
Similar results for a comparable Knudsen number are shown in~\cite{fp-comparison-jun} for a single-species ESFP model, which is why the deviations are attributed to the model itself rather than the mixture modeling.
The results for the rotational temperatures are similar due to the thermal equilibrium in the Couette ﬂow.

\section{Conclusion}
Different continuum-based particle methods, namely an ESBGK model, a Shakhov BGK model, and an ESFP model are proposed.
Aiming for solutions of multi-scale problems, the handling of molecular gas mixture flows is implemented in the open-source particle code PICLas.
Two different supersonic Couette flows with molecular gas mixtures indicate overall very good agreement of the results between both the BGK models and the DSMC method, whereas small deviations occur for the ESFP model.
In future work, the implementation of chemical reactions into all the different models is envisioned.

\section*{Acknowledgments}
This project has received funding from the European Research Council (ERC) under the European Union’s Horizon 2020 research and innovation programme (grant agreement No. 899981 MEDUSA).

%
%

\end{document}